\documentclass[10pt,aps,prl,reprint,nontitlepage,twocolumn,superscriptaddress,floatfix]{revtex4-2}
\usepackage[colorlinks,linkcolor=blue,citecolor=blue,hyperindex,bookmarks=false,pdfstartview=FitH]{hyperref}
\usepackage{gensymb}
\usepackage{mathrsfs}
\usepackage{amsfonts}
\usepackage{amsmath}
\usepackage{amssymb}
\usepackage{graphicx}
\usepackage{color}
\usepackage{subfigure}
\usepackage{microtype}
\usepackage{placeins}
\usepackage{float}
\usepackage{xcolor}
\usepackage{enumitem}
\providecommand{\U}[1]{\protect\rule{.1in}{.1in}}

\providecommand{\U}[1]{\protect\rule{.1in}{.1in}}
\definecolor{myblue}{rgb}{0.00,0.00,0.80}
\definecolor{myred}{rgb}{0.80,0.00,0.00}
\definecolor{mygreen}{rgb}{0.00,0.60,0.00}

\newcommand{\figpanel}[2]{\hyperref[#1]{\ref*{#1}(#2)}}

\begin{document}

\title{Selective high-order topological states and tunable chiral emission in atomic metasurfaces}

\author{Yi-Xin Wang}
\affiliation{School of Physics and Center for Quantum Sciences, Northeast Normal University, Changchun 130024, China}
\author{Yan Zhang}
\email{zhangy345@nenu.edu.cn}
\affiliation{School of Physics and Center for Quantum Sciences, Northeast Normal University, Changchun 130024, China}
\author{Lei Du}
\email{lei.du@chalmers.se}
\affiliation{Department of Microtechnology and Nanoscience (MC2), Chalmers University of Technology, 412 96 Gothenburg, Sweden}
\author{Lingzhen Guo}
\affiliation{Center for Joint Quantum Studies and Department of Physics,
School of Science, Tianjin University, Tianjin 300072, China}
\author{Jin-Hui Wu}
\email{jhwu@nenu.edu.cn}
\affiliation{School of Physics and Center for Quantum Sciences, Northeast Normal University, Changchun 130024, China}

\begin{abstract}
Atomic metasurfaces (AMs) provide a powerful nanophotonic platform for integrating topological effects into quantum many-body systems. 
In this Letter, we investigate the quantum optical and topological properties of a two-dimensional Kagome AM, going beyond the tight-binding approximation and incorporating all-to-all interactions. 
We reveal selective higher-order topological states with a unique dynamical ``chasing" behavior, protected by a generalized chiral symmetry and enabling efficient topological directional transfer.
By introducing an impurity atom---a giant atom---coupled to all array atoms, we observe chiral emission patterns strongly dependent on the atomic polarization.
This nonlocal coupling structure allows exploration of self-interference effects at subwavelength scales. 
Our findings establish AMs as a versatile platform for engineering tunable topological states and chiral quantum optical phenomena, with potential applications in customized light sources and photonic devices.
\end{abstract}

\maketitle

\emph{Introduction.---}Subwavelength atomic arrays (SAAs) have emerged as a versatile paradigm for engineering exotic light-matter interactions~\cite{Mad1, PRL2021Emitter, PRA2022Dirac, PRL134.123602} and advancing topological quantum optics~\cite{Top1}, enabled by many-body dipole-dipole interactions and collective effects such as superradiance and subradiance. 
This quantum metamaterial supports applications in quantum networks~\cite{Qet2, Qet4, PRL2024MLEntanglement} and cooperative optical devices~\cite{Pem1, Pem4, Pem5}. 
Two-dimensional (2D) SAAs---termed atomic metasurfaces (AMs)---enable intriguing functionalities including dipole blockade~\cite{PhBlo1, PhBlo2}, magnetic mirrors~\cite{OMR1, Pem2}, and chiral sensing~\cite{PRL2024ChiralSening}, while also extending to domains such as 2D quantum electrodynamics~\cite{2DAQED} and optomechanics~\cite{opm2}. 
Furthermore, they present significant prospects in  coherent light source~\cite{PRL2020lightsource}, quantum battery~\cite{PRA2025quantumbattery}, chirality-induced  spin-orbit coupling~\cite{PRA2024SOcoupling}, atomic-cavity QED~\cite{PRA2025ACQED}, and quantum computing~\cite{PRA2024computing}.
Their ability to host long-lived collective states enables coherent photon storage~\cite{Scs2, Qme3} and the realization of integer~\cite{Top2, Top5} and fractional quantum Hall effects~\cite{Top3, FraHall1, FraHall2}, underscoring their promise in topological photonics.

Higher-order topological states have garnered significant interest across diverse platforms, including topological metal~\cite{Jin2023255},  photonic~\cite{HToPho1, HToPho2, HToPho4}, electric~\cite{HToEle1}, microwave~\cite{HToMicW2}, acoustic~\cite{HToPhono2,zheng2025}, and atomic~\cite{HToA2} systems.
These systems enable promising functionalities, such as  photon trapping~\cite{HTA2}, topological lasing~\cite{SciAdv2023Hlaser}, and high-fidelity adiabatic quantum state transfer~\cite{PRL2024ToPumo}. 
However, most studies rely on tight-binding models~\cite{TopBook}, limiting the accessible topological features. 
AMs offer a promising route to overcome such constraints, facilitating the realization of higher-order topology in quantum optical settings.

\begin{figure}[hb]
\centering
\includegraphics[width=0.45\textwidth]{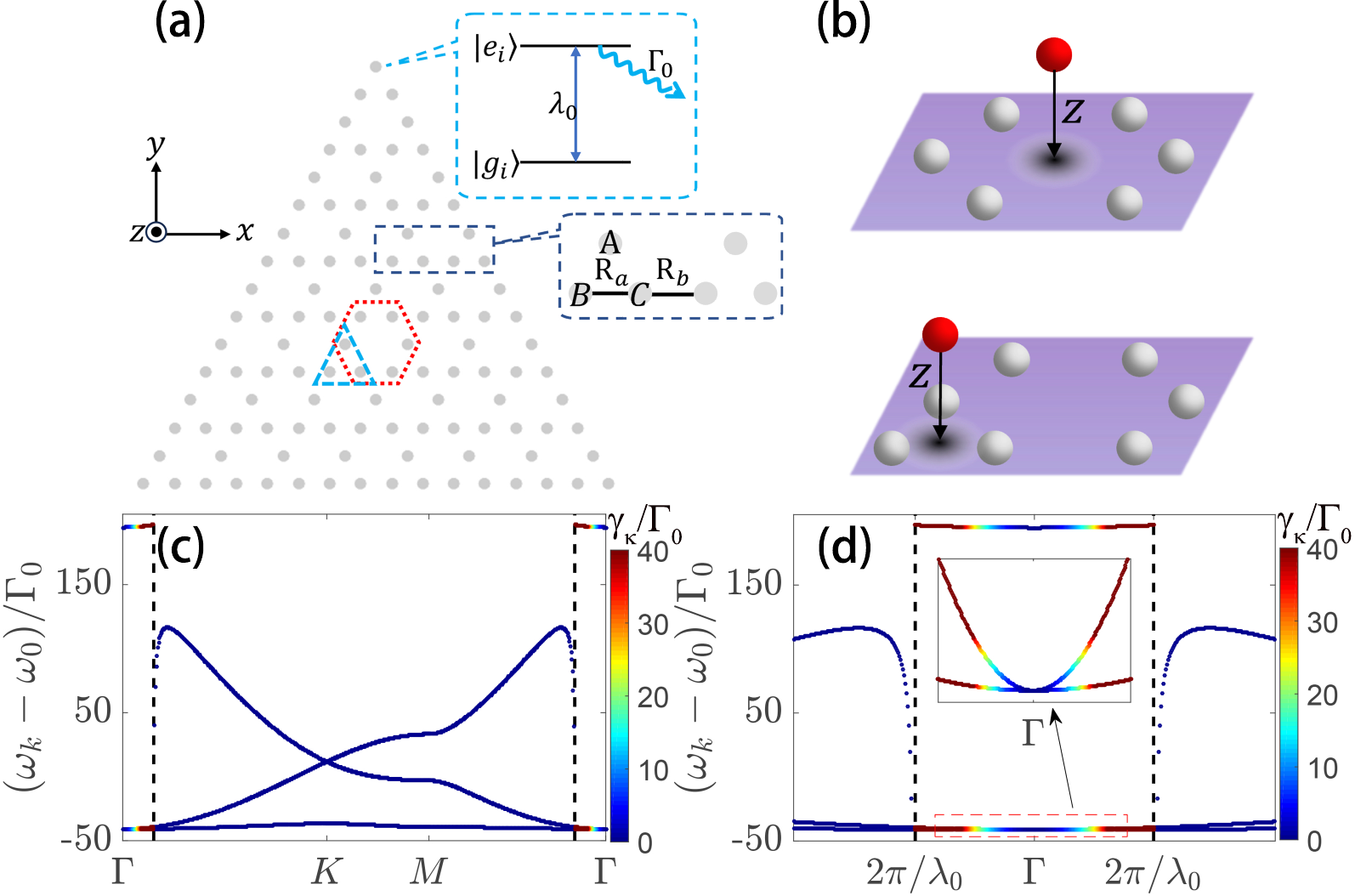}
\caption{(a) Schematic of an AM, where a unit cell consists of sites $A$, $B$, and $C$
Red (blue) dashed lines mark the central hexagon plaquette (adjacent cell).
(b)Two cases of the positions of the impurity atom above the AM.  
(c) Energy band structure for the array.
(d) Band structure near the light cone; the insert magnifies the region around the crossing. 
Black dashed lines represent the light cone,i.e., the Bloch wave vector $|\boldsymbol{k }|=\sqrt{k_x^2+k_y^2}=2\pi/\lambda_0$ with the $x$-axis and $y$-axis components $k_{x,y}$.
Here $\hat{\boldsymbol{\wp}}_0=\hat{\mathbf{e}}_z$, $\lambda_0=790\,$nm, $\Gamma_0=2\pi\times6\,$MHz, $d=0.1\lambda_0$, $z=0.4d$, $\Gamma_A=0.002\Gamma_0$ and $\delta=0$.}
\label{Fig.1}
\end{figure}

In parallel, another emerging quantum paradigm, dubbed ``giant atoms (GAs)", which couple to a field at multiple separate points, has challenged the conventional notion of local light-matter interactions~\cite{TransmissionLine2017, WANGPRA2020, kockum2021}. These systems exhibit self-interference effects that give rise to phenomena such as frequency-dependent relaxation and Lamb shifts~\cite{GALamb.PRA2014}, in-band decoherence-free interactions~\cite{GASupConLamb.Nat2020,complexDFI,Qgate2}, oscillating bound states~\cite{GuoPRR2020}, controllable frequency conversion~\cite{DLFreConvers.PRA2021, DLFreConvers.PRR2021}, and chiral BICs and scattering~\cite{BICPRL2021,GAWG1}, to name a few. GAs have been realized in superconducting circuits~\cite{Science2014, SupquancirGA5, GuoNonMar.PRA2017, YouJQNC}, and explored in platforms such as photonic synthetic dimensions~\cite{Syn2DL}, modulated optical lattices~\cite{PRL2019GAtom}, and Rydberg atoms~\cite{ChenYTprr}.  
However, GA effects at subwavelength scales and the role of atomic polarization remain largely unexplored. This positions AMs as a compelling platform for future investigation.

In this Letter, we study a 2D Kagome AM, going \emph{beyond the tight-binding approximation} by incorporating far-field, all-to-all interactions that are essential at subwavelength scales. 
We focus on two key aspects beyond the scope of simplified tight-binding models: (i) the emergence of higher-order topological characteristics, which are modulated by the orientation of atomic polarization, and (ii) the extension of GA effects to subwavelength regimes, revealing how atomic polarization shapes the directional emission patterns of an impurity GA.

\emph{Kagome atomic metasurfaces.---}We consider a 2D Kagome AM with intracell spacing $R_a$ and intercell spacing $d=R_a+R_b$, where $R_a=(1+\delta)d/2$ defines a spacing imbalance $\delta$ [see Fig.~\figpanel{Fig.1}{a}]. 
By using the dipole approximation and adiabatically eliminating the photonic degrees of freedom under the Born-Markov approximation, the effective non-Hermitian Hamiltonian governing atom-atom interactions in the single-excitation manifold~\cite{Mad1, Top1} is
\begin{align}
  H_B &= \hbar\sum_{i=1}^{N}\left( \omega_0-i\frac{\Gamma_0}{2} \right)\sigma_{ee}^{i} \nonumber\\
  &\quad +\frac{3\pi\hbar c \Gamma_0}{\omega_0}\sum_{i\neq j} \left[ \hat{\boldsymbol{\wp}}_i^{*}\cdot \mathbf{G}_{ij}\cdot \hat{\boldsymbol{\wp}}_j \right] \sigma_{eg}^{i}\sigma_{ge}^{j},
  \label{array-H}
\end{align}
where $\hat{\boldsymbol{\wp}}_i=\boldsymbol{\wp}_i/|\boldsymbol{\wp}_i|$ is the polarization unit vector of the $i$th atom with dipole moment $\boldsymbol{\wp}_j$, $\mathbf{G}$ is the dyadic Green's function (see Sec.~S1 of~\cite{SM}), $\Gamma_0$ is the free-space radiative linewidth, $N$ is the total number of atoms in the array, $\sigma_{ab}^{i}=|a_i\rangle\langle b_i|$ $(a,b=e,g)$ are atomic spin operators, and $c$ is the speed of light in vacuum. Here, we consider all atoms to be identical two-level systems, with transition wavelength (frequency) $\lambda_0$ ($\omega_0$) and the polarization aligned along a common orientation $\hat{\boldsymbol{\wp}}_0$.

Under periodic boundary conditions, the single-excitation eigenmodes take the Bloch form
\begin{align}
  |\psi_{\mathbf{k}}\rangle=\sum_n e^{i\mathbf{k}\cdot\mathbf{R}_n} \left[ p_{\mathbf{k}}^A|e_A^n\rangle+p_{\mathbf{k}}^B|e_B^n\rangle+p_{\mathbf{k}}^C|e_C^n\rangle \right],
  \label{Bloch-wavefunction}
\end{align}
with the Bloch wave vector $\mathbf{k}$ and the center position $\mathbf{R}_n$ of each unit cell.
The non-Hermitian eigenfunction $H_B|\psi_{\mathbf{k}}\rangle=\hbar(\omega_{\mathbf{k}}-i\gamma_{\mathbf{k}})|\psi_{\mathbf{k}}\rangle$ can be solved using the Green’s function regularization (see Sec.~S2 of~\cite{SM}), where $\omega_{\mathbf{k}}$ ($\gamma_{\mathbf{k}}$) is the energy dispersion (decay rate) of the AM. 

Compared to simplified tight-binding models (see Sec.~S3 of~\cite{SM}), retarded long-range interactions substantially reshape the energy band structure of the Kagome AM. 
$\delta=0$ (i.e., $R_a=R_b$) marks a topological phase transition; $\delta > 0$ ($\delta < 0$) corresponds to the topological nontrivial (trivial) phase (see Sec.~S5 of~\cite{SM}). 
Across most of the first Brillouin zone, the decay rate is minimized, enabling longer photon lifetimes and thereby enhanced propagation through the AM. 
Superradiant modes are confined within the light cone region, as shown in Fig.~\figpanel{Fig.1}{d}. 
At the $\Gamma$ point, which is located at the center of the Brillouin zone inside the light cone, the decay rate vanishes since dipole radiation is forbidden when aligned with $\hat{\boldsymbol{\wp}}_0$, resulting in subradiant modes. 
Similarly, modes near the $\Gamma$ point also exhibit suppressed decay rates ($\gamma_{\mathbf{k}}<\Gamma_0/2$, see Fig.~\figpanel{Fig.1}{d}). 
Strong coupling to free space induces superradiant modes near or inside the light cone, whereas beyond it, momentum mismatch leads to subradiant behavior.
The dispersion $\omega_{\mathbf{k}}$ exhibits saddle points at the $M$ points in both the upper and middle bands, giving rise to van Hove singularities.

\begin{figure}[tpb]
\begin{center}
\includegraphics[scale=0.5]{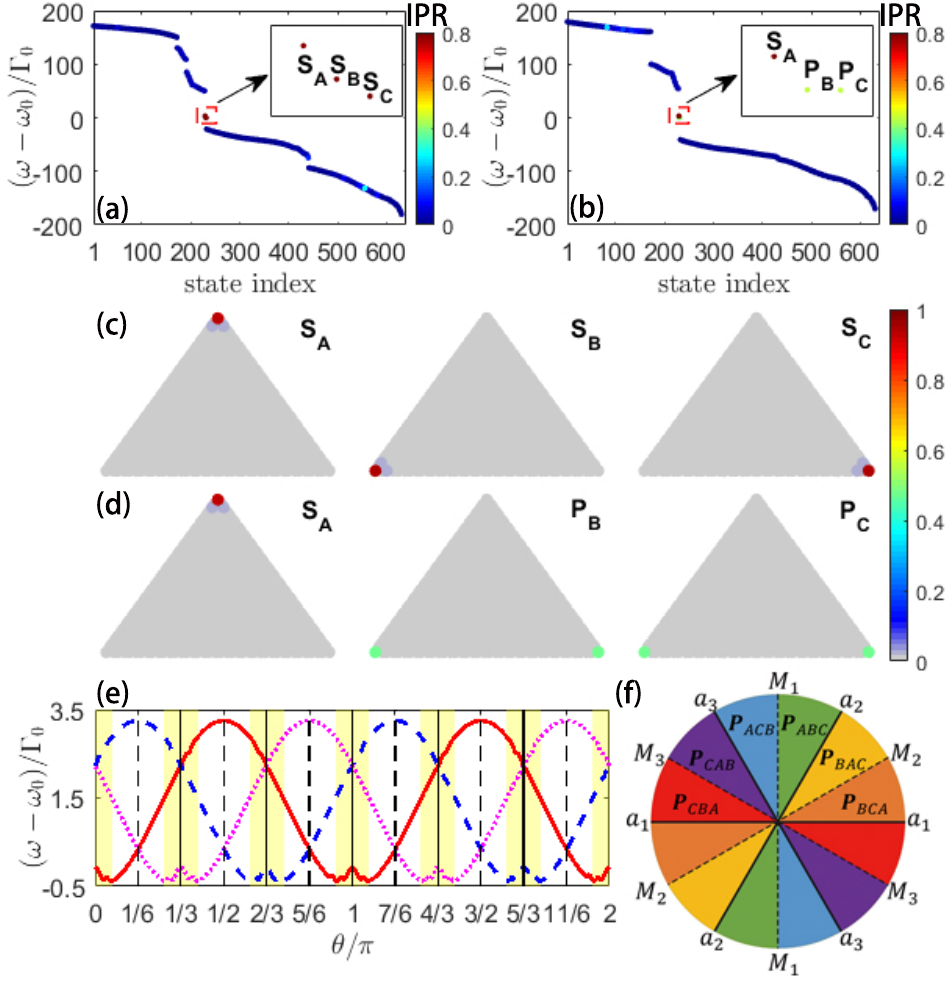}
\caption{Spectra of the AM versus state index with angles (a) $\theta=4\pi/9$ and (b) $\theta=\pi/2$, color-coded by the inverse participation ratio (IPR) $\sum_{j=1}^{N}|p_j|^4/(\sum_{j=1}^{N}|p_j|^2)^2$.
Inserts magnify corner modes. 
(c) and (d) Population distributions $|p_j|^2$ correspond to (a) and (b), respectively.
(e) Eigenfrequencies of the corner modes $\mathbf{S}_{A}$ (red solid), $\mathbf{S}_{B}$ (blue dashed), and $\mathbf{S}_{C}$ (magenta dotted) versus $\theta$, with shaded regions indicating the bulk. 
(f) Energy diagram of the three corner modes as a function of $\theta$, where the sectors $P_{abc}$ ($a,b,c=A, B, C$) indicate the ordering $\omega_{\mathbf{S}_a}>\omega_{\mathbf{S}_b}>\omega_{\mathbf{S}_c}$. 
Solid (dashed) boundaries correspond to directions $M_i$ ($a_i$) ($i=1,2,3$), where $a_{1,2,3}$ denote the directions of three edges of the triangular array ($\theta=m\pi+0, \pi/3, 2\pi/3$).  
The polarization $\hat{\boldsymbol{\wp}}_0$ is in-plane.
Here $\delta=0.3$, and other parameters are the same as those in Fig.~\ref{Fig.1}.}
\label{Fig.2}
\end{center}
\end{figure}

\emph{Selective topological higher-order states.---}
We first explore the topological properties of the AM, focusing on exotic higher-order topological states protected by the intrinsic symmetry of the array. 
Based on all-to-all interactions, the atomic polarization orientation serves as an additional degree of freedom, which enables the emergence of novel higher-order states beyond the reach of tight-binding models~\cite{Tve1, Tve2, KHT, GCS2019}.

\begin{figure}[ptb]
\begin{center}
\includegraphics[scale=0.5]{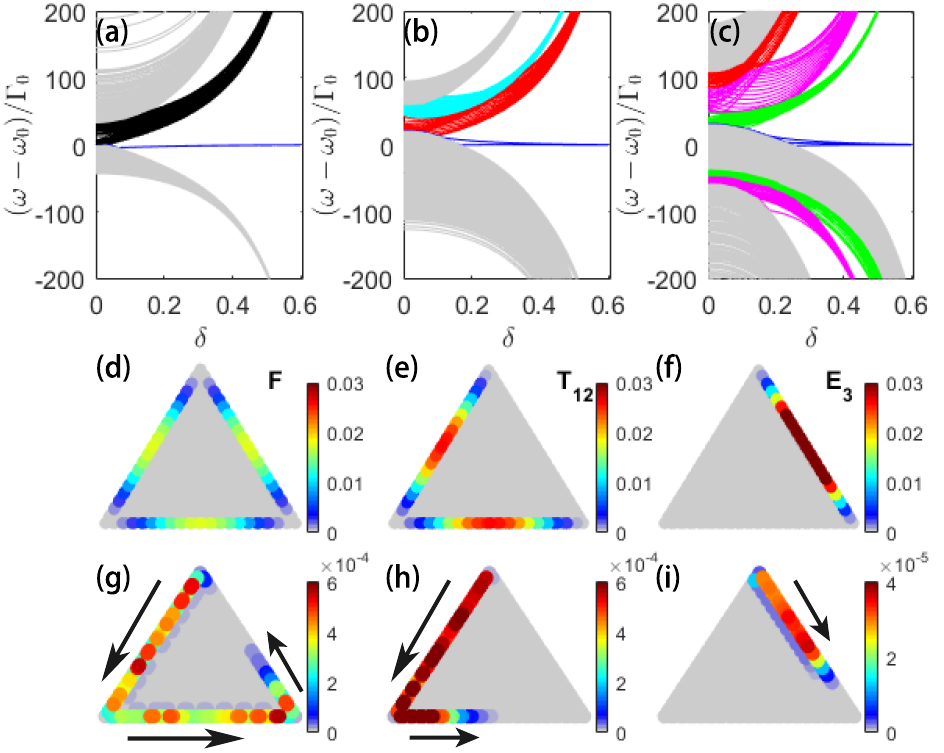}
\caption{Spectra of the AM versus spacing imbalance $\delta$ for (a) out-of-plane polarization $\hat{\boldsymbol{\wp}}_0=\hat{\mathbf{e}}_z$, (b) in-plane polarization $\theta=\pi/6$, and (c) in-plane polarization $\theta=\pi/4$. 
Corner modes (blue) remain separated from the bulk band (gray).
Full-edge $\mathbf{F}$ (black), double-edge $\mathbf{T}_{12}$ (cyan), and single-edge modes $\mathbf{E}_{1}$ (green), $\mathbf{E}_{2}$ (magenta), and $\mathbf{E}_{3}$ (red) are shown.
Population distributions of edge states $\mathbf{F}$, $\mathbf{T}_{12}$, and $\mathbf{E}_{3}$ are displayed in (d-f), respectively. Snapshots of the time evolution of the edge states with $\omega_A-\omega_0=64.3\Gamma_0$, (g) at $t\Gamma_0=2$ for $\hat{\boldsymbol{\wp}}_A=\hat{\mathbf{e}}_z$, (h) at $t\Gamma_0=3.3$ for $\hat{\boldsymbol{\wp}}_A=-\hat{\mathbf{e}}_x-i\hat{\mathbf{e}}_y$, and (i) at $t\Gamma_0=3.3$ for $\hat{\boldsymbol{\wp}}_A=\hat{\mathbf{e}}_x-i\hat{\mathbf{e}}_y$. 
Here $\delta=0.6$, and other parameters are the same as those in Fig.~\ref{Fig.1}.}
\label{Fig.3}
\end{center}
\end{figure}

In the topologically nontrivial phase ($\delta>0$), the bulk polarization is nonzero (see Sec.~S5 of~\cite{SM}), giving rise to localized states at specific boundary sites.
For out-of-plane polarizations ($\hat{\boldsymbol{\wp}}_0=\pm\hat{\mathbf{e}}_z$), the array supports topological corner states protected by the $C_{3v}$ point group. 
These states manifest as either full-corner states or dipole-like double-corner states~\cite{SMHPo1} on the triangular AM (see Sec.~S6 of~\cite{SM}). This is in contrast to the tight-binding model, which supports only full-corner states.
For in-plane polarizations ($\hat{\boldsymbol{\wp}}_{0}=\cos\theta\hat{\mathbf{e}}_x+\sin\theta\hat{\mathbf{e}}_y$), the point-group symmetry reduces to $C_{2v}$ when $\hat{\boldsymbol{\wp}}_0$ aligns with one of the high-symmetry axes $M_i$ (corner bisectors), occurring at polarization angles $\theta = m\pi+\pi/2,\pi/6, 5\pi/6$ ($m = 0, \pm1, \pm2, \ldots$) for $M_{1,2,3}$, respectively.
For all other values of $\theta$, the symmetry reduces to $C_s$, and the polarization orientation breaks the degeneracy of the three topological modes $\mathbf{S}_A$, $\mathbf{S}_B$, and $\mathbf{S}_C$ within the band gap [see Fig.~\figpanel{Fig.2}{a}]. 
As a result, three single-corner modes emerge [see Fig.~\figpanel{Fig.2}{c}].

However, when certain spatial symmetry is restored, double-corner states may emerge [see Fig.~\figpanel{Fig.2}{d}]. For example, by aligning $\hat{\boldsymbol{\wp}}_0$ along $M_1$ at \(\theta = \pi/2\) [see Fig.~\figpanel{Fig.2}{b}], the single-corner modes \(\mathbf{S}_B\) and \(\mathbf{S}_C\) reorganize into a dipolar-like double-corner state $\mathbf{P}_B$ and a monopolar-like double-corner state $\mathbf{P}_C$.
Two double-corner modes are slightly non-degenerate with $\omega_{\mathbf{P}_{B}}-\omega_{\mathbf{P}_{C}} \simeq 0.04\Gamma_0$, arising from the long-range interactions within the same sublattice. 
Owing to the generalized chiral symmetry~\cite{GCS2019}, the corner modes remain predominantly localized on their respective sublattices.

Moreover, the eigenfrequencies of these topological modes exhibit a periodic dependence on the polarization angle $\theta$ [see Fig.~\figpanel{Fig.2}{e}], with a relative phase difference $\pm\pi/3$ between them. 
This leads to a ``chasing'' behavior between the energies of the three corner modes, where the modes interchange their energy ordering at specific $\theta$ values. 
For instance, at $\theta=\pi/15$, the frequency ordering is given by $\omega_{\mathbf{P}_{B}}>\omega_{\mathbf{P}_{C}}>\omega_{\mathbf{P}_{A}}$. 
As $\theta$ increases, $\mathbf{S}_{A}$ overtakes $\mathbf{S}_{C}$ and gradually approaches $\mathbf{S}_{B}$. 
At specific $\theta$ values marked by black-dashed lines in Figs.~\figpanel{Fig.2}{e}, two of the modes among $\mathbf{S}_{A, B, C }$ reorganize into $\mathbf{P}_{A, B, C}$ modes near the detuning value of $0.339\Gamma_0$.
Despite merging into the bulk in certain $\theta$ regions, the corner states retain strong localization and continue the ``chasing" behavior due to the generalized chiral symmetry.
This behavior is illustrated in the phase diagram in Fig.~\figpanel{Fig.2}{f}, which shows the sequential emergence of the three corner states as the polarization angle is tuned.
The chasing behavior enables exotic modulation of higher-order topological BICs~\cite{Topo-BIC-PRB2019, Topo-BIC-PRB2020} (see Sec.~S7 of~\cite{SM}). 
This capability lays the foundation for designing topologically reconfigurable quantum routers, repeaters, memory units, adaptive network configurations, and error correction codes for quantum computing and networking applications.

Figures~\figpanel{Fig.3}{a}--\figpanel{Fig.3}{c} depict the eigenvalue spectra for $C_{3v}$, $C_{2v}$, and $C_{s}$ symmetries, respectively. They reveal that the corner modes remain pinned near a common eigenvalue, exhibiting only slight spectral shifts, while the edge modes persist.
As $\delta$ increases, both the corner and edge modes gradually separate from the bulk bands, leading to enhanced localization at the corners and edges, respectively.
Figures~\figpanel{Fig.3}{d}--\figpanel{Fig.3}{f} further show that the spatial distributions of the edge states are governed by their respective point-group symmetries. 
Notably, beyond the conventional full-edge states achievable in simplified models, the present system allows for tunable double- or single-edge states with controllable spatial positioning.

More interestingly, the dynamics of these edge modes may inspire the development of topological waveguide QED.  
As an example, we consider an emitter with frequency $\omega_A$ and polarization $\hat{\boldsymbol{\wp}}_A$, placed above the top site of the array.  
Under $C_{3v}$ symmetry, the array supports \emph{bidirectional} guided modes, resembling those in conventional structured waveguides, when driven by an excited atom undergoing a $\pi$-transition. 
This behavior arises from the time-reversal symmetry of the atom-field interaction [see Fig.~S8(b) of~\cite{SM}].  
However, \emph{unidirectional} guided modes can be achieved via chiral emission, when the array is excited by circularly polarized light [see Fig.~\figpanel{Fig.3}{g}], due to the chirality of the near fields associated with the array modes~\cite{AQED1, AQED2, AQED3, TAW-QED2023PRR}.
Such directional transport can also be customized based on the double- and single-edge states, as shown in Figs.~\figpanel{Fig.3}{h} and~\figpanel{Fig.3}{i}, where the field propagates along specific boundaries of the AM while remaining isolated from others. 
This mechanism does not rely on the polarization of the emitter; instead, it only requires engineering the spatial distributions of the double- and single-edge states.
Such edge states, unachievable in simplified models, enable the formation of topologically protected atomic waveguides along specific edges (see Sec.~S8 of~\cite{SM}). This further underscores the versatility of the AM in engineering structured reservoirs.

\begin{figure}[ptb]
\begin{center}
\includegraphics[width=0.48\textwidth]{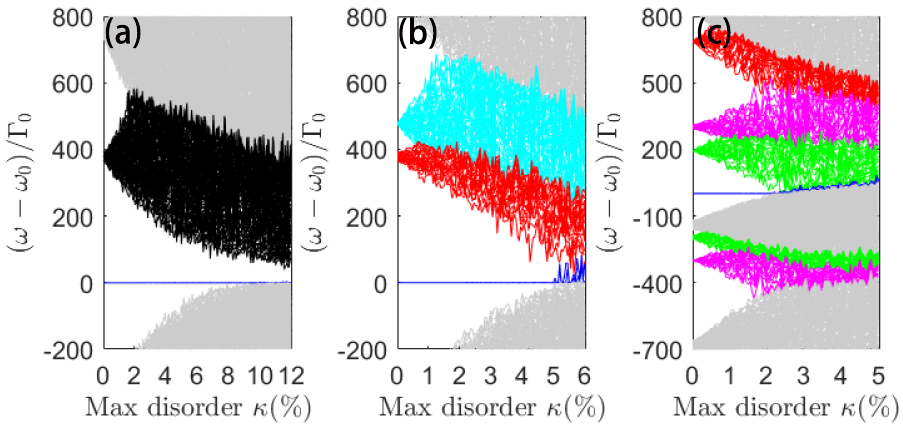}
\caption{Spectra of the AM with increasing positional disorder $\kappa$ for (a) out-of-plane polarization $\hat{\boldsymbol{\wp}}_0=\hat{\mathbf{e}}_z$,  (b) in-plane polarization $\theta=\pi/6$, and (c) in-plane polarization $\theta=\pi/4$.
Colors follow those in Fig.~\ref{Fig.2}. 
Here $\delta=0.6$, and other parameters are the same as those in Fig.~\ref{Fig.1}.}
\label{Fig.4}
\end{center}
\end{figure}

\emph{Robustness against disorder.---}We now examine the robustness of topological modes against positional disorder by introducing a random shift $\kappa=|\boldsymbol{\kappa}|$ to each atomic position, i.e., $\mathbf{r}\rightarrow\mathbf{r}+\boldsymbol{\kappa}R_a$. As shown in Figs.~\figpanel{Fig.4}{a}--\figpanel{Fig.4}{c}, the disorder breaks the symmetry of the AM, lifting the ($2+1$)-fold degeneracy of the corner modes in the out-of-plane polarization case and effectively reducing the point group to $C_s$ in the in-plane polarization case. As a result, the monopolar- and dipolar-like modes merge into the single-corner states. While the edge modes gradually blend into the bulk, the corner modes remain well isolated within the band gap.
Thus, the initial symmetry of the system plays a key role in determining its robustness. 
It shows from Fig.~\ref{Fig.4} that the corner modes remain robust up to disorder strengths of $\kappa\simeq 9.7\%$, $ 5.2\%$, and $2.3\%$ for $C_{3v}$, $C_{2v}$, and $C_{s}$ symmetries, respectively. 
Such robustness makes them promising for robust photonic or hybrid devices, where tolerance to fabrication imperfections and external disturbances is essential.

\emph{Tunable chiral emission.---}Finally, we reveal intriguing spontaneous emission phenomena in this Kagome AM at the topological transition point. 
To this end, we introduce an impurity atom above the array. 
As shown in Fig.~\figpanel{Fig.1}{b}, the impurity atom---with polarization orientation $\hat{\boldsymbol{\wp}}_A$---is positioned above the center of either the central hexagon plaquette (upper panel) or an adjacent unit cell of the central plaquette (lower panel), tuned to resonance with the van Hove singularity. 
Note that this impurity atom is \emph{nonlocally} coupled to the entire array at the subwavelength scale, effectively forming a GA. 
In this case, the total Hamiltonian is given by $H=H_B+H_A+H_I$, where
\begin{eqnarray}
  H_A&=&\hbar\left( \omega_A-i\frac{\Gamma_A}{2}\right)\sigma_{ee}^A,
  \label{impurity-H}\\
  H_{I}&=&\frac{3\pi c\hbar\sqrt{\Gamma_0\Gamma_A}}{\omega_0}\sum_m^N \left\{ \left[ \hat{\boldsymbol{\wp}}_A^{*}\cdot \mathbf{G}_{Am}\cdot \hat{\boldsymbol{\wp}}_m \right]\sigma_{eg}^{A}\sigma_{ge}^{m} \right.\nonumber\\
&&\left.+\left[ \hat{\boldsymbol{\wp}}_m^{*}\cdot \mathbf{G}_{mA}\cdot \hat{\boldsymbol{\wp}}_A \right]\sigma_{eg}^{m}\sigma_{ge}^{A}\right\}.
  \label{impurity-array-interaction-H}
\end{eqnarray}
The impurity atom can be either identical to the array atoms or a different isotope, with its transition frequency and linewidth tunable via Raman-assisted processes~\cite{PRL2022DE}.
Here, provided that $\Gamma_A\ll\Gamma_0$~\cite{AQED1, PRL2022DE}, the AM evolves on a much faster timescale than the impurity. This allows the AM to serve as a Markovian structured bath.

\begin{figure}[t]
\centering
\includegraphics[width=0.48\textwidth]{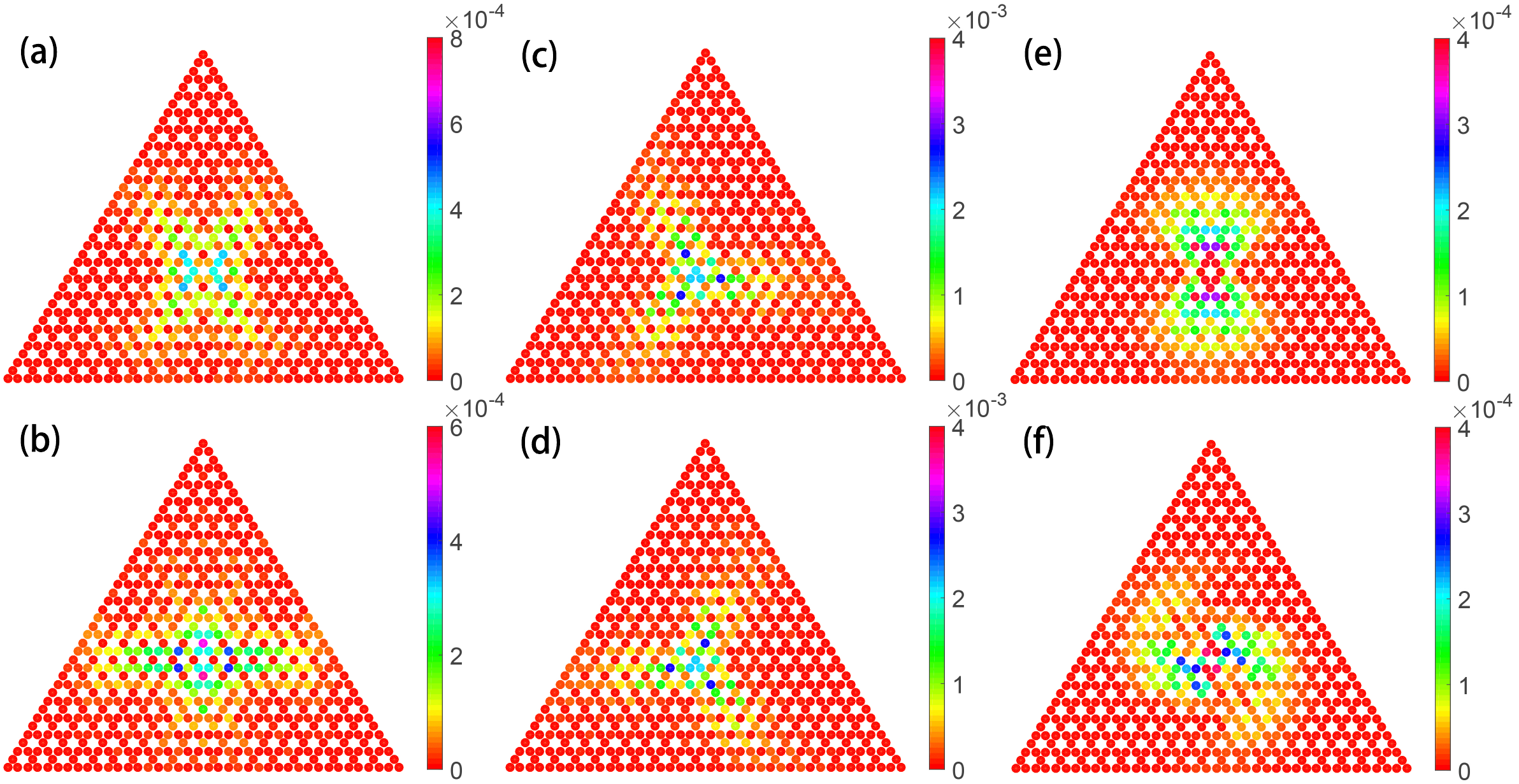}
\caption{Population distributions of the AM. 
(a) and (b) Distributions at $t\Gamma_0=0.3$, for a two-level impurity atom of $\omega_A-\omega_0=-3.06\Gamma_0$, with (a) $\hat{\boldsymbol{\wp}}_A=\hat{\mathbf{e}}_x$ and (b) $\hat{\boldsymbol{\wp}}_A=\hat{\mathbf{e}}_y$.
(c) and (d) Distributions at $t\Gamma_0=0.45$, for a two-level impurity atom of $\omega_A-\omega_0=-3.06\Gamma_0$, with (c) $\hat{\boldsymbol{\wp}}_A=-\hat{\mathbf{e}}_x-i\hat{\mathbf{e}}_y$ and (d) $\hat{\boldsymbol{\wp}}_A=\hat{\mathbf{e}}_x-i\hat{\mathbf{e}}_y$. 
Distributions at $t\Gamma_0=0.17$, for a three-level impurity atom of $\omega_A-\omega_0=48.26\Gamma_0$, with (e) $B=0$ and (f) $B=20\Gamma_0$. 
In (a), (b), (e), and (f) [(c) and (d)], the position of the impurity atom corresponds to the upper (lower) panel in Fig.~\figpanel{Fig.1}{b}.
Other parameters are the same as those in Fig.~\ref{Fig.1}.
}
\label{Fig.5}
\end{figure}

For $\hat{\boldsymbol{\wp}}_A=\hat{\mathbf{e}}_x$, the two-level impurity GA exhibits a symmetric cross-emission pattern along two sides of the triangular Kagome array (not aligned with the $x$-axis), consistent with tight-binding predictions (see Sec.~S3 of~\cite{SM}) [see Fig.~\figpanel{Fig.5}{a}]. 
For $\hat{\boldsymbol{\wp}}_A=\hat{\mathbf{e}}_y$, the emission predominantly follows the $\pm x$-axis in the subradiant regime [see Fig.~\figpanel{Fig.5}{b}].
Remarkably, the emission exhibits a threefold chiral symmetry, with the preferential directions separated by $\pi/3$ [see Figs.~\figpanel{Fig.5}{c} and \figpanel{Fig.5}{d}], when the impurity GA undergoes $\sigma^{\mp}$-transitions.
These findings highlight that the emission direction of the impurity atom is highly sensitive to both its polarization orientation and spatial position. 
This tunability opens new avenues for engineering exotic emission patterns and topologically protected chiral quantum networks~\cite{lodahl2017chiral}.

To explore more exotic emission patterns, we now extend our investigation to a three-level $V$-type impurity GA, offering additional internal degrees of freedom and richer emission dynamics. 
Specifically, the GA features a ground state $|g\rangle$ and two excited hyperfine states $|\sigma_{+,-}\rangle$.
Two transitions are associated with $\hat{\boldsymbol{\wp}}_A=\mp \hat{\mathbf{e}}_x-i\hat{\mathbf{e}}_y$, respectively.
The atomic Hamiltonian is given by $H_{A}^{V}=\hbar\sum_{\alpha=+,-}\left( \omega_A+\alpha\mu B-i\frac{\Gamma_0}{2} \right) \sigma_{\alpha}^{\dag}\sigma_{\alpha}$, where $\mu B$ is the Zeeman shift of the hyperfine states with magnetic moment $\mu$, induced by a magnetic field $\mathbf{B}=B\hat{\mathbf{e}}_z$. 
For the initial state $|\psi(0)\rangle=(|\sigma_{+}\rangle+|\sigma_{-}\rangle)/\sqrt{2}$, the GA exhibits a longitudinal emission pattern when $B=0$ [see Fig.~\figpanel{Fig.5}{e}], consistent with tight-binding predictions (see Sec.~S3 of~\cite{SM}). 
When $B\neq 0$, the lifting of the degeneracy of the excited states leads to distinct resonant couplings near the van Hove singularity. 
This creates \emph{spiral-like} emission patterns [see Fig.~\figpanel{Fig.5}{f}] and extended lifetimes, enabled by subradiant modes near the singularities [see Fig.~\figpanel{Fig.1}{c}]. 
These chiral emission behaviors may enable applications in chiral quantum dynamics~\cite{PRL2024CQD, Nat2017chiral}, chiral transport~\cite{PRX2018CT}, and superradiant bursts~\cite{PRX2024SB}.

\emph{Conclusions.---}In summary, we have proposed a triangular Kagome-structured AM with all-to-all interactions, and explored its higher-order topological states and chiral transport behaviors---combined with a GA---in the topological regime, beyond the tight-binding approximation. 
By using spatial point group variations induced by the polarization of the array atoms, we demonstrate symmetry-dependent single- and double-corner (edge) states with selective spatial positioning. 
Notably, the single-corner states exhibit a ``chasing" behavior, dynamically transitioning between the band gap and the continuum while remaining localized as BICs. 
Additionally, these selective single- or double-edge states enable directionally guided modes, which provide a promising platform for engineering topologically protected waveguides.  
This mechanism enables reconfigurable and selective localization of quantum states.
The robustness of these topological modes and their polarization-controlled dynamics facilitate chiral quantum transport~\cite{SM}, offering new avenues for atomic-waveguide QED and robust long-range dipole-dipole interactions~\cite{SCI2018_359}.
At the topological phase transition, we have further examined chiral emission from an impurity GA coupled to all atoms of the AM, revealing polarization-dependent emission behaviors and subwavelength effects. 
These results fill the knowledge gap of GA effects at subwavelength scales.
Harnessing all-to-all interactions and polarization responses opens opportunities for customizing photonic environments and tunable light-matter interfaces.
Our work lays the foundation for the development of topological photonics and the advancement of chiral quantum networks.

This work is supported by the Jilin Scientific and Technological Development Program (Grant No. 20240101328JC), the Knut och Alice Wallenberg stiftelse (Grant No. 2022.0090), the Science Foundation of the Education Department of Jilin Province (Grant No. JJKH20250301KJ), and the National Natural Science Foundation of China (Grant No. 62375047)

\end{document}